\documentclass[pra,preprint,showpacs]{revtex4}

\usepackage{bm}
\usepackage{amsmath}
\usepackage{dcolumn}
\usepackage{soul}
\usepackage{xcolor}

\begin{document}

\title{Absolute numbering of asymptotic vibrational levels of diatomic molecules from cold physics experiments}

\author{A. Pashov} 
\email{pashov@phys.uni-sofia.bg}
\affiliation{Faculty of Physics, Sofia University, 5 James
Bourchier blvd., 1164 Sofia, Bulgaria}
\author{P. Kowalczyk} 
\email{Pawel.Kowalczyk@fuw.edu.pl} 
\affiliation{Institute of Experimental Physics, Faculty of
Physics, University of Warsaw, ul. Pasteura 5, 02-093 Warsaw,
Poland} 
\author{W. Jastrzebski} 
\email{jastr@ifpan.edu.pl}
\affiliation{Institute of Physics, Polish Academy of Sciences,
Al.Lotnik\'{o}w 32/46, 02-668~Warsaw, Poland}

\date{\today}

\begin{abstract}
We present a simple method for determination of absolute vibrational numbering of isolated near dissociation levels in diatomic molecules, usually observed in cold physics experiments. The method is based on the isotope shift and works even when energies of only two levels from one isotopologue and one level from another isotopologue have been measured. It is demonstrated on data from recently reported precise measurements of binding energies of levels lying close to the dissociation limits in ultracold Yb$_2$, CsYb, RbSr and RbYb molecules. Its predictions agree with these of much more elaborate multi-isotope potential curve fitting.
\end{abstract}


\maketitle

\section{Introduction}

In a series of recently published papers, measurements of binding energies for the few last rovibrational levels in the ground electronic state were reported in RbYb \cite{RbYb}, Yb$_2$ \cite{Yb2}, CsYb \cite{CsYb} and RbSr \cite{RbSr} molecules. These studies were focused on the scattering properties of the corresponding ultra-cold atomic species, including the interatomic interaction potentials and scattering lengths. The experiments were carried out at very low temperatures in atomic traps which limited the characterization of the ground states to their near dissociation region. Nevertheless in each of the studies, by using advanced multi-isotope models based on potential curve fit, the authors were able to account for the common phase of the near asymptotic wave functions which is due to the motion of the nuclei in the corresponding potential curve at short and intermediate internuclear distances. As a result the ground state potential energy curves were constructed and the total number of vibrational levels supported by the potential (i.e. the absolute vibrational numbering of the levels) was determined for various isotopologues. Such information is essential e.g. for prediction of positions and widths of Feshbach resonances, calculation of scattering lengths for various isotopic combinations etc. 

In our recent paper \cite{isonum} we discussed the problem of finding the absolute vibrational numbering in an electronic state and showed that the proper treatment of multi-isotope data may allow to get it even from a fragmentary data set. The main idea in \cite{isonum} is quite general and relies only on a smooth dependence of the rovibrational energy on quantum numbers $v$ and $J$. The particular approach applied there, however, was not directly applicable for levels close to the dissociation limit. The experimental data reported in papers \cite{RbYb,Yb2,CsYb,RbSr} motivated us to develop further the ideas of Ref.~\cite{isonum} and to check their validity for near asymptotic levels, thus offering an additional, purely algebraic tool to establish the true vibrational numbering in an electronic state. As already mentioned in Ref.~\cite{isonum} such a simple approach should not be considered as an alternative to the potential curve multi-isotope analysis, but rather as a supplementary tool which allows for a fast determination of the absolute vibrational numbering from few raw spectroscopic data. This numbering will put a serious constraint on the initial potential curves fitted in the more sophisticated multi-isotope analysis and can reduce the computational efforts, even when the uncertainty in the established absolute value of $v$ exceeds one vibrational quantum. If the last vibrational level is observed experimentally, the analysis allows to fix the total number of vibrational levels and to test the accuracy of theoretical potentials obtained beforehand by quantum chemistry methods.

From a fundamental point of view, the presented algebraic approach stresses the fact that both local vibrational spacing and the isotope shift depend rather on the overall depth of the potential energy curve and not on its particular shape. This property might be easily overlooked/hidden in the complicated multi-isotope fits of potential curves or molecular constants.

\section{Background}

In Ref.~\cite{isonum} it was discussed that within the single channel approximation, the energy levels of a diatomic molecule may be expressed as a function $F(x,y)$ of the reduced vibrational and the rotational quantum numbers $x=\alpha(v+0.5)$ and $y=\alpha^2J(J+1)$. For a given isotopologue of the molecule with reduced mass $\mu_0$ we assume $\alpha=1$. Reduced masses $\mu_{\mathrm{iso}}$ of other isotopologues are accounted for through $\alpha=\sqrt{\mu_0/\mu_{\mathrm{iso}}}$. The fact that the energy levels of all isotopologues can be described by the same function makes it possible to find the absolute vibrational numbering even in fragmentary data set, by fitting positions of rovibrational levels of all isotopologues simultaneously, either using the Dunham expansion or a potential energy curve.

It was demonstrated \cite{isonum} that if the $F(x,y)$ function is known even for narrow interval of energies, and one knows the isotopic shift for just one energy level within this interval, it is possible to determine the absolute vibrational numbering of that level. In brief, let $E_0=F(x_0,y_0)$ be the energy of the main isotopologue and $E_{\mathrm{iso}}=F(x_{\mathrm{iso}},y_{\mathrm{iso}})$ the energy of the other one. Both levels correspond to the same absolute vibrational quantum number. Since the function $F$ is known and single valued it is possible to invert it and write

\begin{equation}
x_0=G(E_0,y_0) \mbox{\hspace{10mm} and \hspace{10mm}} x_{\mathrm{iso}}=G(E_{\mathrm{iso}},y_{\mathrm{iso}})
\end{equation}

Since $x_0=(v+0.5)$ and $x_{\mathrm{iso}}=\alpha(v+0.5)$, the vibrational number of the pair is

\begin{equation}
v= \frac{x_{\mathrm{iso}}-x_0}{\alpha-1}-0.5=\frac{G(E_{\mathrm{iso}},y_{\mathrm{iso}})-G(E_0,y_0)}{\alpha-1}-0.5   \mbox{ .}
\label{main}
\end{equation}

In \cite{isonum} it was shown that the shape of $F(x,y)$ (and also $G(E,y$)) may be approximated locally by fitting only few pairs of spectral lines which makes this approach very attractive for determination of the absolute vibrational numbering in case of a sparse experimental data set. It is very often the case when photoassociation data are considered. Unfortunately, the approximate expressions for $F(x,y)$ suggested in Ref.~\cite{isonum} are not applicable near the dissociation limit of the molecule. Here we demonstrate that the method can be extended also to this region by applying the Near Dissociation Expansion (NDE) to fit the shape of $F(x,y)$. NDE is an expression for the energies of the last few levels in a potential well with asymptotic $D_e-C_m/R^m$ behaviour \cite{NDE}

\begin{equation}
E_v=D_e-a(v_D-v)^{\frac{2m}{m-2}} \mbox{ .}
\label{ndeexp0}
\end{equation}

\noindent where $D_e$ is the energy of the atomic asymptote, $v_D$ is the noninteger vibrational index at the dissociation limit and $a$ is a parameter depending on $C_m$ and its power $m$ \cite{NDE}. The same expansion can be written using the reduced quantum number $x$

\begin{equation}
E_v=D_e-a(x_D-x)^{\frac{2m}{m-2}} \mbox{ .}
\label{ndeexp}
\end{equation}

\noindent Here $x_D=v_D+0.5$. Inverting eq.~(\ref{ndeexp}) we obtain

\begin{equation}
x=x_D-\Big(\frac{D_e-E}{a}\Big)^{1/k} \mbox{ ,}
\label{ndex}
\end{equation}

\noindent where notation

\begin{equation}
k=\frac{2m}{m-2}
\end{equation}

\noindent is used. Note that in all cases discussed below we deal with ground molecular states in which the long range behaviour is governed by the van der Waals interaction, therefore $m=6$ and $k=3$.

The expression (\ref{ndeexp}) is based on the simplest form of the NDE with single dispersion term $C_m$ and no rotational energy. We will show, however, that even in this form the expression correctly predicts results for the cases of interest. In Ref.~\cite{NDE2} an improved formula for the NDE has been suggested. For the simple asymptotic potential with one dispersion coefficient it has been written in the form:

\begin{equation}
v_D-v=H_m^{-1}(D_e-E)^{\frac{m-2}{2m}}+\gamma(D_e-E)
\label{Comparat}
\end{equation}

\noindent  where the parameter $H_m^{-1}$ depends on the dispersion coefficient $C_m$ and $\gamma$ is a correction coefficient, both defined analytically  (eqs. (14) and (15) in \cite{NDE2}). It can be noticed that (\ref{Comparat}) actually describes the rotationless inverse function $G(E,y)$. The more general form of the improved formula (eq. (35) in \cite{NDE2}) is given again as an inverse function, which can be directly used for determination of the absolute vibrational numbering in place of eq.~(\ref{ndex}).

For more complex forms of NDE (e.g. discussed in Ref.~\cite{NDE1}) derivation of an analytic expression for the inverse function $G(E,y)$, replacing eq.~(\ref{ndex}), may be not straightforward or even possible. Nevertheless if the function $F(x,y)$ and energy $E$ are known, one can always find $x$ by solving the equation $E=F(x,y)$ numerically.

The calculated vibrational numbers $v$ should be matched to integers and this may lead to some ambiguity, especially when they are close to a half integer. In order to provide quantifiable metric that allows unbiased assessment of the assigned vibrational numbers, it is necessary to compare the calculated vibrational numbers $v$ for several observed levels. Let us have $M$ experimental levels for the minor isotopologues $E_{\mathrm{iso}}$. From (\ref{ndeexp}) we calculate $M$ values $v_i$ and assign to each of them an integer $v_i^{\mathrm int}$. We will use the average deviation:

\begin{equation}
\Delta v_{\mathrm{ave}}=\frac{1}{M}\sum_{i=1}^M(v_i-v_i^{\mathrm int})
\end{equation}

\noindent to find integer values for which $\Delta v_{\mathrm{ave}}$ is closest to $0$.

The proposed method works only if the energies $E_0$ and $E_{\mathrm{iso}}$ belong to levels with the same absolute vibrational numbers. Matching such pairs in the experimental data may be also sometimes ambiguous. As shown in \cite{isonum} (see eq.13 there), matching the level $E_0$ to a level $E_{\mathrm{iso}}$ for which the assumed vibrational number differs from the correct one by one quantum will lead to a change of the calculated vibrational numbers by $\pm\alpha/(\alpha-1)$. For heavier molecules, like those considered here, $\alpha=\sqrt{\mu_0/\mu_{\mathrm{iso}}}$ is very close to 1, and therefore the shift of the vibrational numbering will be huge, which usually can be excluded by comparison even with crude theoretical calculations.

In the following sections we present separately the application of formula~(\ref{main}) for determination of vibrational numbering in the ground electronic states of several molecules (RbYb \cite{RbYb}, Yb$_2$ \cite{Yb2}, CsYb \cite{CsYb} and RbSr \cite{RbSr}), by assuming the simplest form of the NDE, given by eq.~(\ref{ndeexp}).

\section{Y\lowercase{b}$_2$}

In \cite{Yb2} Borkowski and coauthors reported on a sub-kHz photassociation spectroscopy of three isotopologues of Yb$_2$. Energies of the last three vibrational levels in the ground state $0^+_g$ (for $J=0$ and $2$) have been determined with respect to the dissociation limit. These data were described within the experimental uncertainty by two multi-isotope models including potential energy curves. In one of the models the Born-Oppenheimer breakdown effects also have been accounted for. One of the important outcomes of the modelling is that the ground state potential energy curve of $^{174}$Yb$_2$ contains 72 vibrational levels. This results from a simultaneous fit of all experimental data containing 13 energy levels of $^{168}$Yb$_2$, $^{170}$Yb$_2$ and $^{174}$Yb$_2$ with a single potential energy curve. The limited data set and the large extrapolation towards the minimum of the potential energy curve may raise questions on the validity of this result, but in fact it is based on the same assumptions as we are doing in our method \cite{isonum} and therefore both approaches should lead to the same correct number of vibrational states. The difference between our analysis and that of Ref.~\cite{Yb2} is that the first one is much simpler, whereas the second one provides a potential energy curve with correct near asymptotic behaviour and the vibrational numbering comes as a consequence. 

We shall discuss the case of Yb$_2$ molecule in more details than the other ones to make our presentation more instructive. In Table~\ref{Bork} we display experimental results of Borkowski \emph{et al.} concerning binding energies of all observed levels with $J=0$, retaining the relative vibrational numbering $p$ of the original work. Note however, that Borkowski \emph{et al.} denoted relative vibrational quantum number for the least bound level as $p=1$ and increased the numbers when going down from the atomic asymptote. But to account properly for the isotope shift, the numbering of vibrational levels should start from the bottom of the potential well, so that vibrational quantum numbers increase by approaching the asymptote. Therefore we just changed the sign of the relative vibrational quantum numbers from \cite{Yb2}. We shall use label $n$ hereafter for the proper relative vibrational quantum numbers. As one can see from Table~\ref{Bork}, for $^{174}$Yb$_2$ $n=-p$. However, this is not the case for the other isotopologues, because the numbering used in Ref.~\cite{Yb2} was introduced independently for each of them. The same potential curve may support a different number of levels for isotopologues of different masses, for instance the least bound levels with $p=1$ in $^{170}$Yb$_2$ and $p=1$ in $^{174}$Yb$_2$ do not necessarily correspond to the same absolute quantum number $v$ (and to the same relative number $n$). Indeed, we shall argue that it is not so. The heavier the isotope, the higher should be the binding energy of a level with given $v$ and $n$. Therefore the $p=1$ level in $^{170}$Yb$_2$ with binding energy $E_b=-27.700$~MHz cannot correspond to $p=1$ level in $^{174}$Yb$_2$ with $E_b=-10.625$~MHz, but rather to the next level $p=2$ with $E_b=-325.664$~MHz. By the same argument the $p=2$ level in $^{168}$Yb$_2$ corresponds to $p=2$ in $^{170}$Yb$_2$ and $p=3$ in $^{174}$Yb$_2$. Note that this leaves the $p=1$ level in $^{174}$Yb$_2$ without a counterpart, which means either that in $^{174}$Yb$_2$ the potential well contains one level more than in the other isotopologues, or that some of the levels labelled with $p=1$ are not the least bound ones (which is quite unlikely according to \cite{Yb2}). The last column in Table~\ref{Bork} contains the relative quantum numbers $n$ established in the way described above.

\begin{table}
\caption{Binding energies for vibrational levels of three different isotopologues of Yb$_2$ measured in~\cite{Yb2} and their relative numbering. The numbers in parentheses indicate uncertainty of energy in last quoted digits.}
\centering
\begin{tabular}{c c c c }
\\
isotopologue & relative number $p$ & binding energy & relative label $n$ \\
 & from Ref.~\cite{Yb2} & [MHz] & (our analysis) \\
\hline
&&&\\
 $^{168}$Yb$_2$ & $2$ & $-195.18141(46)$ & $-3$ \\
 $^{170}$Yb$_2$ & $1$ & $-27.70024(42)$ & $-2$ \\
				& $2$ & $-463.72552(8)$ & $-3$ \\
				& $3$ & $-1922.01467(505)$ & $-4$ \\
 $^{174}$Yb$_2$ & $1$ & $ -10.62513(53)$ & $-1$ \\
				& $2$ & $-325.66378(98)$ & $-2$ \\
				& $3$ & $-1527.88543(34)$ & $-3$ \\ 
				&&&\\
\hline
\end{tabular}
\label{Bork}
\end{table}

In summary, for $^{168}$Yb$_2$ one level with $n=-3$ is recorded (hereafter we use our relative numbering), for $^{170}$Yb$_2$ -- three levels with $-4\leq n\leq -2$ and for $^{174}$Yb$_2$ also three levels  with $-3\leq n\leq -1$. Since for $^{174}$Yb$_2$ the vibrational levels closest to the dissociation limit have been observed, we took it as the reference isotopologue and fitted parameters of the NDE formula (\ref{ndeexp}) (note that for the main isotopologue the reduced vibrational quantum number $x$ in the formula equals to $n+0.5$), obtaining $a= 99.9306(4)$ MHz and $x_D= -0.018020(4)$ ($D_e=0$ by definition). These parameters define locally the shape of the $F(x,y)$ function as well as its inverse (\ref{ndex}).

As an example, in order to obtain the absolute vibrational number $v$ of the level with $E_{\mathrm{iso}}=-195.18141(46)$ MHz in $^{168}$Yb$_2$ (labelled with $n=-3$, see the first row of Table~\ref{results}), first we identify the corresponding level in the reference isotopologue $^{174}$Yb$_2$ -- it corresponds also to $n=-3$ and its energy is $E_0=-1527.88543(34)$ MHz. Then we need the reduced quantum numbers for both levels. For the $^{174}$Yb$_2$ level it is $x_0=n+0.5=-2.5$. For the $^{168}$Yb$_2$ level the reduced quantum number can be calculated from formula~(\ref{ndex}) using the fitted NDE parameters which provides $x_{\mathrm{iso}}\approx-1.268$. With the known value of $\alpha$ (also listed in Table~\ref{results} for $^{168}$Yb$_2$), substitution of all values to formula~(\ref{main}) yields:

\begin{equation}
v= \frac{x_{\mathrm{iso}}-x_0}{\alpha-1}-0.5=\frac{-1.268+2.5}{1.017722-1}-0.5 \approx 69.018   \mbox{ .}
\end{equation}

By the same method absolute vibrational numbers $v$ can be calculated for all pairs of levels corresponding to different isotopologues (one of them should be always the reference isotopologue, $^{174}$Yb$_2$) and having the same relative number $n$. From the results listed in Table~\ref{results} one can see that the label $n=-3$ corresponds to $v=69$ and $n=-2$ to $v=70$. So for the least bound level (with label $n=-1$) the absolute vibrational quantum number is $v=71$, in excellent agreement with the analysis of \cite{Yb2}. Given the uncertainty of the experimental binding energies, the uncertainty of $v$ might be estimated by following the propagation of errors through the expressions (\ref{ndeexp}) and (\ref{main}). However, the uncertainty of $0.0003$ vibrational quanta obtained in this way (see Table~\ref{results}) is unrealistically low. The reason is that the simple NDE model (\ref{ndeexp}) cannot fit the experimental data to within the sub-kHz accuracy. As a consequence the uncertainties of NDE parameters and all related quantities are underestimated. The actual difference between the experimental and the calculated NDE energies are on average about 0.3 MHz which is roughly a factor 1000 larger than the experimental uncertainty. So the model applied here would be adequate even in case of the experimental uncertainty around 0.3 MHz and than the uncertainty of $v$ would be $0.2$ -- $0.3$, which is a more realistic estimate for the uncertainty of the absolute vibrational numbers.

The agreement between the calculated vibrational numbers and the corresponding integer values is very good. The average deviation based on three experimental points is $\Delta v_{\mathrm{ave}}=0.13$. One can argue, however, that the relative vibrational numbers from Table~\ref{Bork} are somehow arbitrary. Why not, for example, matching $p=1$ of $^{170}$Yb$_2$ with $p=3$ of $^{174}$Yb$_2$. This would mean that $p=1$ for $^{170}$Yb$_2$ would become $n=-3$ and $p=2$ of the same isotopologue $n=-4$. Repeating the calculations with the new assignment we obtain $v=155.7\pm 0.0003$ for $n=-3$ and $v=154.5\pm 0.0003$ for $n=-4$ (the shift from the previous assignment, leading to $v=70$ and $v=69$, is consistent with $\alpha/(\alpha-1)\approx 86.4$ \cite{isonum}). Given the experimental uncertainty, these values deviate from integer values more than at the previous assignment: $\Delta v_{\mathrm{ave}}=0.4$. To further check to consistency of this assignment we consider also the measurements for the isotopologue $^{168}$Yb$_2$. We cannot leave $p=1$ to correspond to the level assigned as $n=-3$ because then its calculated vibrational number $v=69.0$ contradicts the numbering for $^{170}$Yb$_2$. If we match $p=1$ to $n=-4$, then the calculated vibrational number becomes $v=125.4\pm 0.0002$ which is again far from an integer and also very far from the value around 155 consistent with $^{170}$Yb$_2$ and $^{174}$Yb$_2$. Obviously, the experimental data cannot lead to consistent numbering with the alternative assignment. But even if data for $^{168}$Yb$_2$ were missing, and we accepted $v=156$ and $v=155$ as absolute vibrational numbers, this would lead to a much deeper potential curve which contradicts the theoretical calculations \cite{Yb2_abinitio} (see also eq. (13) from \cite{isonum} and the discussion there on the mismatch of the vibrational numbers).

\section{C\lowercase{s}Y\lowercase{b}}

Guttridge \emph{et al.} \cite{CsYb} performed two-photon photoassociation spectroscopy of three isotopologues $^{133}$Cs$^{170}$Yb, $^{133}$Cs$^{173}$Yb, $^{133}$Cs$^{174}$Yb and measured binding energies of a few highest vibrational levels in the ground X$^2\Sigma^+_{1/2}$ state, including the least bound level. They determined the absolute vibrational numbering of the observed levels by fitting a properly parametrized theoretical potential curve of the X state to the binding energies of levels in all three isotopologues, concluding that the ground state potential supports 77 vibrational levels.

Since the least bound level is labelled in Ref.~\cite{CsYb} with relative vibrational quantum number $n=-1$ and the labels decrease for stronger bound levels, we adopted the same numbering here. The experimental uncertainty of binding energies reported in Table I of Ref.~\cite{CsYb} varies from 0.1 MHz to 2 MHz. We used the three least bound levels for Cs$^{174}$Yb to find the NDE parameters ($a= 93.47(6)$ MHz, $x_D  = 0.4442(4)$). Then by applying the same procedure as for Yb$_2$ the absolute vibrational numbering was established (see Table~\ref{results}). Due to the larger experimental uncertainty the ambiguity in $v$ is larger than for Yb$_2$. At first glance the least bound level should correspond to $v=77$, but this would be in contradiction with the vibrational numbering of the next two levels (those labelled with $n=-2$ and $n=-3$). Indeed, let us find the average deviation $\Delta v_{\mathrm{ave}}$ between the calculated vibrational numbers and the corresponding integers. If $n=-1$ corresponds to $v=77$, then $n=-2$ corresponds to $v=76$ and $n=-3$ to $v=75$. From Table~\ref{results} we can see that the deviations between the calculated $v$ for levels with $n=-2$ and $n=-3$ and these integer values are about one vibrational quantum and the average deviation is $\Delta v_{\mathrm{ave}}=0.76$. If however $n=-1$ corresponds to $v=76$, only one calculated $v$ (the one for $n=-1$ in Cs$^{173}$Yb with $v=77.1$) deviates by more than 1, while the average deviation for all five data is $\Delta v_{\mathrm{ave}}=-0.24$ -- a much better agreement. So $v=76$ is a much more reasonable choice for the label $n=-1$ which is also the result of the more intricate analysis in \cite{CsYb}.

\section{R\lowercase{b}S\lowercase{r}}

Ciamei \emph{et al.} \cite{RbSr} performed two-colour photoassociation spectroscopy of  $^{87}$Rb$^{84}$Sr, $^{87}$Rb$^{87}$Sr and $^{87}$Rb$^{88}$Sr molecules. By combining the ultracold data with spectroscopic results obtained at hot environment they constructed an isotopically consistent potential energy curve of the ground $^2\Sigma^+$ state, although this required bridging an energy gap of about 900~cm$^{-1}$ where the experimental data were missing. The authors predict 67 bound levels for $^{87}$Rb$^{84}$Sr and $^{87}$Rb$^{87}$Sr and 68 bound levels for $^{87}$Rb$^{88}$Sr.

We took the experimental data from Table II of Ref.~\cite{RbSr}. The highest vibrational level in  $^{87}$Rb$^{84}$Sr and $^{87}$Rb$^{87}$Sr was predicted in Ref.~\cite{RbSr} to be $v=66$.  Arbitrarily we labelled it with $n=3$ and $v=65$ with $n=2$. For our analysis we used only levels with $N=0$ and $F=1$. From the data for $^{87}$Rb$^{84}$Sr we determined the NDE parameters: $a=215.01(80)$ MHz and $x_D=4.013(2)$. By adding data concerning the other isotopologues three pairs of levels with the same vibrational label could be selected. Their absolute vibrational numbers were found using formula (\ref{main}) and they are listed in Table~\ref{results}. The experimental accuracy for RbSr molecule is comparable to that for CsYb \cite{CsYb}, so the uncertainties of the vibrational numbers are also comparable. Again the present simple approach is in agreement with the analysis from Ref.~\cite{RbSr}. 

\section{R\lowercase{b}Y\lowercase{b}}

Borkowski \emph{et al.} \cite{RbYb} report on spectroscopic investigation of the highest vibrational levels in the ground electronic state $^2\Sigma_{1/2}$ of $^{87}$RbYb molecule containing four different isotopes of ytterbium, done by two-colour photoassociation. The provided data (Table I in \cite{RbYb}) are the most abundant, but also the least precise. The experimental error is estimated to be about 7 MHz for the most accurate measurement and may reach several hundred MHz for others. Several model potentials were constructed by the authors which reasonably reproduce the experimental observations but differ in well depth, thus imposing different vibrational numbering on the levels. The potentials support between 61 and 70 bound levels with the curve containing 66 levels being the authors' choice.

Proceeding with our analysis, one should be careful with the vibrational labels used in Ref.~\cite{RbYb}. Similarly to the Yb$_2$ case discussed earlier, the original labels start from 1 for the least bound level and increase for levels with higher binding energy. For isotopologues containing $^{174}$Yb and heavier ytterbium isotopes, the ground state potential well contains one level more than for the lighter ones \cite{PhDMuenchow}. This means that for example the levels labelled with $1$ for $^{87}$Rb$^{170}$Yb and $^{87}$Rb$^{176}$Yb in fact correspond to different absolute vibrational numbers. Since our analysis relies on energies of levels corresponding to the same absolute vibrational number, it was necessary to change the labels adopted in \cite{RbSr}. Similarly to the Yb$_2$ case we changed signs of the original labels and in addition increased the labels for $^{87}$Rb$^{176}$Yb by one. Thus the experimental level with $E=-304$ MHz labelled originally with $2$, in the present study became $n=-1$ since it has the same absolute vibrational number as the level with energy $E=-113.3$ MHz in $^{87}$Rb$^{170}$Yb. 

The isotopologue $^{87}$Rb$^{176}$Yb was selected as the reference one since the experimental data are the most accurate for it and the derived NDE parameters were the most certain ($a= 150.7(32)$ MHz, $x_D=0.764(20)$). 

The results for RbYb molecule displayed in Table~\ref{results} show that the experimental uncertainty may be not sufficient to establish reliably the absolute vibrational numbering, the uncertainty on $v$ amounts to several vibrational quanta. However, it should be borne in mind that the authors of Ref.~\cite{RbYb} had also difficulties to determine the number of levels in the ground state potential and they have chosen somewhat arbitrarily 66 levels for $^{87}$Rb$^{176}$Yb with an uncertainty of several vibrational quanta. This would mean that the label $n=-1$ (the last but one vibrational level) corresponds to $v=64$ and $n=-2$ to $v=63$. Within the stated uncertainties, the results of Table~\ref{results} are not in contradiction with the multi-isotope analysis in \cite{RbYb}, but still indecisive because of insufficient accuracy of the measurements.

\begin{table}
\fontsize{8pt}{13pt}\selectfont
\caption{Absolute vibrational quantum numbers $v$ of levels deduced from the experimental data on Yb$_2$ \cite{Yb2}, CsYb \cite{CsYb}, RbSr \cite{RbSr} and RbYb \cite{RbYb}, using formulas (\ref{ndex}) and (\ref{main}). $E$ is the binding energy in MHz (with uncertainty in last quoted digits given in parentheses), $n$ -- the relative vibrational number, $x$ -- the reduced vibrational quantum number and  $\alpha$ -- the isotope ratio $\sqrt{\mu_0/\mu_{\mathrm{iso}}}$. See text for more explanations.}
\centering
\begin{tabular}{c l c l c c c}
&&&&&\\
molecule& $E$ [MHz] & $n$ & $x$& $v$ & isotope & $\alpha$ \\
\hline
Yb$_2$ &&&&&&\\
&$-195.18141(46)$ & $-3$ & $-1.268$ & $69.0 \pm 0.0002$ & $^{168}$Yb$_2$ & $1.017722$\\
&$ -27.70024(44)$ & $-2$ & $-0.670$ & $70.4 \pm 0.0003$ & $^{170}$Yb$_2$ & $1.011713$\\
&$-463.72552(80)$ & $-3$ & $-1.686$ & $69.0 \pm 0.0003$ & $^{170}$Yb$_2$ & $1.011713$\\
&$-10.62513(53)	$& $-1$	& $-0.5$&  & $^{174}$Yb$_2$ & $1.000000$ \\
&$-325.66378(98)	$& $-2$	& $-1.5$&  & $^{174}$Yb$_2$ & $1.000000$ \\
&$-1527.88543(34)$& $-3$	& $-2.5$&  & $^{174}$Yb$_2$ & $1.000000$ \\

\hline
CsYb &&&&&&\\
&$  -15.7(3)$ & $-1$ & $-0.108$    &$76.6 \pm  0.4$ & $^{133}$Cs$^{170}$Yb& $1.005090$ \\
&$-1576(2)  $   & $-3$ & $-2.120$  &$74.1 \pm  0.1$ & $^{133}$Cs$^{170}$Yb& $1.005090$\\
&$  -56.8(2)$ & $-1$ & $-0.403$    &$77.1 \pm  0.4$ & $^{133}$Cs$^{173}$Yb& $1.001252$\\
&$ -592(1)  $ & $-2$ & $-1.406$    &$74.5 \pm  0.5$ & $^{133}$Cs$^{173}$Yb& $1.001252$\\
&$-2166(1)  $ & $-3$ & $-2.407$    &$73.9 \pm  0.4$ & $^{133}$Cs$^{173}$Yb& $1.001252$\\
&$-78.7(1)  $& $-1$	& $-0.5$&  & $^{133}$Cs$^{174}$Yb & $1.000000$ \\
&$-686.4(7) $& $-2$	& $-1.5$&  & $^{133}$Cs$^{174}$Yb & $1.000000$ \\
&$-2385.5(9)$& $-3$	& $-2.5$&  & $^{133}$Cs$^{174}$Yb & $1.000000$ \\
\hline
RbSr &&&&&&\\
&$ -29.01(20) $& $3$	& $3.5$&  & $^{87}$Rb$^{84}$Sr & $1.000000$ \\
&$-744.53(20) $& $2$	& $2.5$&  & $^{87}$Rb$^{84}$Sr & $1.000000$ \\
&$ -287.27(20)$ & $3$ & $2.911$ &$66.3 \pm 0.2$ & $^{87}$Rb$^{87}$Sr& $0.991193$\\
&$-1950.24(20)$ & $2$ & $1.927$ &$64.5 \pm 0.2$ & $^{87}$Rb$^{87}$Sr& $0.991193$\\
&$ -458.90(20)$ & $3$ & $2.725$ &$66.2 \pm 0.1$ & $^{87}$Rb$^{88}$Sr& $0.988380$\\
\hline
RbYb &&&&&&\\
&$  -113.3(150) $ & $-1$ & $-0.145$ &$60.4 \pm 3.8$ & $^{87}$Rb$^{170}$Yb& $1.005828$\\
&$ -1028.8(158) $ & $-2$ & $-1.133$ &$62.5 \pm 2.3$ & $^{87}$Rb$^{170}$Yb& $1.005828$\\
&$  -153.8(200) $ & $-1$ & $-0.243$ &$66.4 \pm 6.1$ & $^{87}$Rb$^{172}$Yb& $1.003845$\\
&$ -1237.1(150) $ & $-2$ & $-1.253$ &$63.7 \pm 3.5$ & $^{87}$Rb$^{172}$Yb& $1.003845$\\
&$  -227.9(100) $ & $-1$ & $-0.384$ &$60.5 \pm 7.3$ & $^{87}$Rb$^{174}$Yb& $1.001903$\\
&$ -1477.6(100) $ & $-2$ & $-1.376$ &$64.5 \pm 7.2$ & $^{87}$Rb$^{174}$Yb& $1.001903$\\
&$-304.6(74)^{a)}	   $& $-1$	& $$$-0.5$&  & $^{87}$Rb$^{176}$Yb & $1.000000$ \\
&$-1748.6(88)^{a)}     $& $-2$	& $$$-1.5$&  & $^{87}$Rb$^{176}$Yb & $1.000000$ \\
&$-5256.3(1525)^{a)}   $& $-3$	& $$$-2.5$&  & $^{87}$Rb$^{176}$Yb & $1.000000$ \\
\hline
\multicolumn{4}{c}{$^{a)}$ Average of two experimental values given in \cite{RbYb}}
\end{tabular}
\label{results}
\end{table}

\section{Conclusion}

In this study we extended the method originally presented in Ref.~\cite{isonum} to asymptotic vibrational levels, which were originally excluded from the area of its application. Now we demonstrate the possibility to assess the absolute vibrational numbers from a fragmentary experimental data for levels close to the dissociation limit of a diatomic molecule. The simple algebraic method assumes observation of a single electronic state with smooth dependence of the binding energy on the vibrational and rotational quantum numbers and relies on the measured isotope shifts of spectral lines. 

The obtained results are in agreement with the much more elaborate multi-isotope potential curve fits of the same experimental data \cite{RbYb,Yb2,CsYb,RbSr}. The method presented in this paper may be particularly convenient in case of molecules, for which spectroscopic data on low vibrational levels are missing because of difficulties with experiments in hot environment. Its reliability depends on the accuracy of the experimental data and also on the quality of the model which is used to fit the dependence of the binding energy on $v$ and $J$. For levels far from the dissociation limit a simple Dunham-type expansion proves to be feasible \cite{isonum}. Here we show that close to the atomic asymptote the NDE works equally well. For the molecules discussed here (Yb$_2$, CsYb, RbSr and RbYb) we demonstrated that experimental accuracy of the order of 0.1 --  0.2 MHz is sufficient to reliably establish the absolute vibrational numbers from only few measurements. Higher order effects like the hyperfine structure, breakdown of the Born-Oppenheimer approximation and others, if causing irregular energy shifts below few hundreds kHz, apparently may be neglected for this purpose. 
On the other hand, if the level energies do not depend smoothly on the reduced quantum numbers (for example because of local perturbations larger in magnitude than 0.1 --  0.2 MHz) the proposed method still can help to identify these perturbations.

\section{Acknowledgments}

AP acknowledges partial support from National Science Fund of Bulgaria Grant D18/12/2017. PK and WJ were partially supported by the National Science Centre of Poland (Grant No. 2016/21/B/ST2/02190).

\end{document}